\documentclass[aps,showpacs,floatfix,twocolumn,prl]{revtex4}

\usepackage{epsfig,amsmath,graphicx,graphics,float}

\begin{document}

\title{Quantum Tricriticality and Phase Transitions in Spin-Orbit \\
Coupled Bose-Einstein Condensates}


\author{Yun Li$^{1}$}
\author{Lev P. Pitaevskii$^{1,2}$}
\author{Sandro Stringari$^{1}$}

\affiliation{$^{1}$Dipartimento di Fisica, Universit\`{a} di Trento
and INO-CNR BEC Center, I-38123 Povo, Italy}

\affiliation{$^{2}$Kapitza Institute for Physical Problems, Kosygina
2, 119334 Moscow, Russia}

\begin{abstract}

We consider a spin-orbit coupled configuration of spin-$1/2$
interacting bosons with equal Rashba and Dresselhaus couplings. The
phase diagram of the system at $T=0$ is discussed with special
emphasis on the role of the interaction, treated in the mean-field
approximation. For a critical value of the density and of the Raman
coupling we predict the occurrence of a characteristic tricritical
point separating the spin mixed, the phase separated and the zero
momentum states of the Bose gas. The corresponding quantum phases
are investigated analyzing the momentum distribution, the
longitudinal and transverse spin-polarization and the emergence of
density fringes. The effect of harmonic trapping as well as the role
of the breaking of spin symmetry in the interaction Hamiltonian are
also discussed.

\end{abstract}


\maketitle

A large number of papers have been recently devoted to the
theoretical study of artificial gauge fields in ultracold atomic
gases (for a recent review see, for example, \cite{Dalibard2010}).
First experimental realizations of these novel configurations have
been already become available \cite{Lin2009, Lin2011}. This field of
research looks very promising from both the theoretical and
experimental point of view, due to the possibility of realizing
exotic configurations of non trivial topology \cite{Hasan2010}, with
the emergence of new quantum phases in both bosonic \cite{Wang2010}
and fermionic \cite{Vyasanakere2011, Gong2011} gases, and the
possibility to simulate electronic phenomena of solid state physics.
In the case of Bose gases a key feature of these new systems is the
possibility of revealing Bose-Einstein condensation in
single-particle states with nonzero momentum.


By tuning the Raman coupling between two hyperfine states of
$^{87}$Rb atoms, the authors of \cite{Lin2011} have reported the
first experimental identification of the new quantum phases
exhibited by a spin-orbit coupled Bose-Einstein condensation.
Important features of the resulting phases were anticipated in the
paper by Ho and Zhang \cite{Ho2011} and discussed in the same
experimental paper \cite{Lin2011}. The purpose of this Letter is to
provide a theoretical description of the phase diagram corresponding
to the spin-orbit coupled Hamiltonian employed in \cite{Lin2011}. We
point out the occurrence of an important density dependence in the
phase diagram which shows up in the appearance of a tricritical
point that, to our knowledge, has never been predicted for such
systems.

We will consider the mean-field energy functional (for simplicity we
set $\hbar=m=1$)
\begin{equation}
\begin{aligned}
E(\psi_a,\psi_b)=& \int d^3r \left[\begin{pmatrix} \psi_a^\ast &
\psi_b^\ast
\end{pmatrix} h_0 \begin{pmatrix}\psi_a \\ \psi_b
\end{pmatrix} + \frac{g_{aa}}{2} |\psi_a|^4  \right.\\
& \left.+ \frac{g_{bb}}{2} |\psi_b|^4+ g_{ab}|\psi_a|^2
|\psi_b|^2\right]
\end{aligned} \label{eq:Ham}
\end{equation}
describing an interacting spin-$1/2$ Bose-Einstein condensate at
$T=0$, where $\psi_a$ and $\psi_b$ are the condensate wave functions
relative to the two spin components interacting with the coupling
constants $g_{ij}=4\pi a_{ij}$, with $a_{ij}$ the corresponding
$s$-wave scattering lengths, and
\begin{equation}
h_0= \frac{1}{2}\left[\left(p_x-k_0 \sigma_z\right)^2+
p_\perp^2\right] + \frac{\Omega}{2} \sigma_x + \frac{\delta}{2}
\sigma_z +V_{\text{ext}} \label{eq:h_0}
\end{equation}
is the single-particle Hamiltonian characterized by equal
contributions of Rashba \cite{Rashba1984} and Dresselhaus
\cite{Dresselhaus1955} spin-orbit couplings and a uniform magnetic
field in the $x\,$-$\,z$ plane. In Eq.(\ref{eq:h_0}) $\Omega$ is the
Raman coupling constant accounting for the transition between the
two spin states, $k_0$ is the strength associated with the
spin-orbit coupling fixed by the momentum transfer of the two Raman
lasers, $\delta$ fixes the energy difference between the two
single-particle spin states, $\sigma_i$ are the usual $2\times2$
Pauli matrices, while $V_{\text{ext}}$ is the external trapping
potential.

In the first part of the Letter we will consider uniform
configurations, neglecting the effect of the trapping potential
($V_{\text{ext}}=0$) and assume a spin symmetric interaction with
$g_{aa}=g_{bb}\equiv g$ and  $\delta=0$. The effect of asymmetry
will be discussed afterwards. The ground state condensate wave
function will be determined using a variational procedure based on
the following ansatz for the spinor wave function:
\begin{equation}
\begin{pmatrix} \psi_a\\ \psi_b\end{pmatrix}=\sqrt{\frac{N}{V}}
\left[C_1 \begin{pmatrix} \cos\theta \\
-\sin\theta\end{pmatrix}e^{ik_1x} +C_2 \begin{pmatrix} \sin \theta \\
-\cos\theta\end{pmatrix}e^{-ik_1x}\right] \label{eq:psi_ideal}
\end{equation}
where $N$ is the total number of atoms, $V$ is the volume of the
system. For a given value of the average density $n=N/V$, the
variational parameters are then $C_1$, $C_2$, $k_1$ and  $\theta$.
Their values are determined by minimizing the energy (\ref{eq:Ham})
with the normalization constraint $\sum_{i=a,\,b} \int d^3r
|\psi_i|^2=N$ (i.e., $|C_1|^2+|C_2|^2=1$). Minimization with respect
to $\theta$ yields the general relationship $\theta
=\arccos(k_1/k_0)/2$ ($0\leq \theta \leq \pi/4$), fixed by the
single-particle Hamiltonian (\ref{eq:h_0}). Once the other
variational parameters are determined, one can calculate key
physical quantities like, for example, the momentum distribution
accounted for by the parameter $k_1$, the longitudinal and
transverse spin polarization of the gas
\begin{equation}
\langle\sigma_z\rangle = \frac{k_1}{k_0} \left(|C_1|^2- |C_2|^2
\right),\;\;\; \langle\sigma_x\rangle = - \frac{\sqrt{k_0^2
-k_1^2}}{k_0}
\end{equation}
and the density
\begin{equation}
n(x)=n\left[ 1+2 |C_1C_2| \frac{\sqrt{k_0^2-k_1^2}}{k_0} \cos(2k_1
x+\phi)\right] , \label{eq:nx}
\end{equation}
where $\phi$ is the relative phase between $C_1$ and $C_2$. The
ansatz (\ref{eq:psi_ideal}) exactly describes the ground state of
the single-particle Hamiltonian $h_0$ (ideal Bose gas). In this
case, for $\Omega\leq 2 k_0^2$, the energy, as a function of $k_1$,
exhibits two minima located at the values $\pm k_0\sqrt{1-\Omega^2/
4 k_0^4}$ and the ground state is degenerate, the energy being
independent of the actual values of $C_1$ and $C_2$. For $\Omega> 2
k_0^2$ the two minima disappear and all the atoms condense into the
zero momentum state $k_1=0$.

The same ansatz is well suited to discuss the role of interactions.
By inserting
(\ref{eq:psi_ideal}) into (\ref{eq:Ham}), we find that the energy
per particle $\varepsilon=E/N$ takes the form
\begin{equation}
\varepsilon = \frac{k_0^2}{2} -\frac{\Omega}{2 k_0}
\sqrt{k_0^2-k_1^2} -F(\beta) \frac{k_1^2}{2k_0^2} +
G_1\left(1+2\beta\right) \label{eq:varepsilon}
\end{equation}
where we have defined the dimensionless parameter
$\beta=|C_1|^2|C_2|^2$ ($0 \leq \beta \leq 1/4$), and the function
\begin{equation}
F(\beta) = \left(k_0^2-2G_2\right)+4\left(G_1 + 2G_2 \right) \beta
\label{eq:F_beta}
\end{equation}
with the interaction parameters $G_1 =n \left(g+g_{ab}\right)/4$,
$G_2 =n \left(g-g_{ab} \right)/4$. The variational parameters to
minimize the energy are then $k_1$ and $\beta$.

Let us first consider minimization with respect to $k_1$. If $\Omega
>2 F(\beta)$ the energy (\ref{eq:varepsilon}) is an increasing
function of $k_1$ and the minimum takes place at $k_1=0$. If instead
$\Omega <2 F(\beta)$ one finds that $\varepsilon$ is minimized by
the choice
\begin{equation}
k_1(\beta)= k_0 \sqrt{1- \frac{\Omega^2}{4 \left[F(\beta)
\right]^2}} \label{eq:k1_int} \; ,
\end{equation}
which generalizes the ideal gas result $F=k_0^2$. Equations
(\ref{eq:F_beta}) and (\ref{eq:k1_int}) explicitly show that the
momentum distribution is modified by the interactions. We find the
following result for the energy per particle:
\begin{equation}
\varepsilon = -\frac{\Omega^2}{8 F(\beta)}+ G_1+ G_2\left(1-4\beta
\right). \label{eq:varepsilon_beta}
\end{equation}
The ground state of the system can be found by looking for the
minimum of (\ref{eq:varepsilon_beta}) with respect to $\beta$. One
can easily prove that the second order derivative of
(\ref{eq:varepsilon_beta}) with respect to $\beta$ is negative. This
means that the minimum is achieved at the limiting values of
$\beta$. The ground state is then compatible with the three
following phases:

(I) The \emph{spin mixed} or ``\emph{stripe}'' \emph{phase} with
$k_1\neq 0$, $\beta=1/4$ and hence $\langle \sigma_z\rangle =0$. In
this phase the atoms condense in a superposition of two plane wave
states with wave vector $\pm k_1$ and the density (\ref{eq:nx})
exhibits fringes. This configuration is characterized by a
degeneracy associated with the relative phase between the
coefficients $C_1$ and $C_2$ which fixes the actual spatial position
of stripes.

(II) The \emph{separated phase} with $k_1\neq 0$, $\beta=0$ and
hence $\langle \sigma_z\rangle \neq 0$, where the atoms condense
into a single plane wave state with wave vector either $k_1$
($C_2=0$) or $- k_1$ ($C_1=0$), the actual value being determined by
a mechanism of spontaneous spin symmetry breaking.

(III) The \emph{single minimum} or ``\emph{zero momentum}''
\emph{phase} with $k_1=0$ and $\langle \sigma_z\rangle =0$ where the
atoms condense in the zero momentum state. In this phase the gas is
fully polarized along the $x$ direction ($\langle\sigma_x\rangle =
-1$).

We first notice that the spin mixed phase is compatible only with
positive values of the interaction parameter $G_2$, favoring
antiferromagnetic configurations. In fact in the opposite case
$G_2<0$, the first order derivative $\partial \varepsilon /\partial
\beta$ is always positive and the  ground state is always in the
phase separated configuration (II) or in the zero momentum phase
(III).

In the most interesting $G_2>0$ case, the system will be always in
the phase (I) for small values of the Raman coupling constant
$\Omega$. If the condition
\begin{equation}
k_0^2> 4G_2+\frac{4G_2^2}{G_1} \label{eq:trans_12_cond}
\end{equation}
is satisfied, the systems will exhibit a phase transition (I) to
(II) at the frequency
\begin{equation}
\Omega^{\text{(I-II)}} = 2 \left[\left(k_0^2+G_1 \right)
\left(k_0^2-2G_2\right) \frac{2 G_2}{G_1 +2G_2}\right]^{1/2}.
\label{eq:Omega_T_12}
\end{equation}
This generalizes the result derived in \cite{Ho2011}, which
corresponds to the low density (or weak coupling) limit of
(\ref{eq:Omega_T_12}), i.e., $G_1,\, G_2 \ll k_0^2$. The transition
frequency in this limit approaches the density independent value
\begin{equation}
\Omega^{\text{(I-II)}}_{\text{LD}} = 2 k_0^2 \sqrt{2\gamma
/(1+2\gamma)} \label{eq:Omega_T_12_LD}
\end{equation}
where we have introduced the dimensionless interaction parameter
$\gamma =G_2/G_1=(g-g_{ab})/(g+g_{ab})$. By further increasing
$\Omega$, the system will enter the phase (III) at the frequency
\begin{equation}
\Omega^{\text{(II-III)}} =2\left( k_0^2-2G_2\right)
\label{eq:Omega_T_23}
\end{equation}
This result, in the limit $G_2 \ll k_0^2$, was also discussed in
\cite{Zhang2011}. If instead the condition (\ref{eq:trans_12_cond})
is not satisfied, the transition will occur directly from the phase
(I) to (III) at the frequency
\begin{equation}
\Omega^{\text{(I-III)}} = 2\left(k_0^2+ G_1 \right)- 2 \left[
\left(k_0^2 +G_1\right) G_1\right]^{1/2} . \label{eq:Omega_T_13}
\end{equation}
In the strong coupling limit $G_1 \gg k_0^2$ (\ref{eq:Omega_T_13})
approaches the constant value $k_0^2$.

The critical point where the phase (II) disappears is fixed [see
Eq.(\ref{eq:trans_12_cond})] by the condition $G_1^{(c)} =
k_0^2/4\gamma\left(1+\gamma\right)$, corresponding to the critical
value
\begin{equation}
n^{(c)} =  k_0^2/\left(2\gamma g\right)
\end{equation}
for the density. If $n < n^{(c)}$, one has two transitions (I-II and
II-III), while if $n > n^{(c)}$, only one phase transition (I-III)
can take place.

\begin{figure}
\centering
\hspace{-1mm}\includegraphics[scale=0.678]{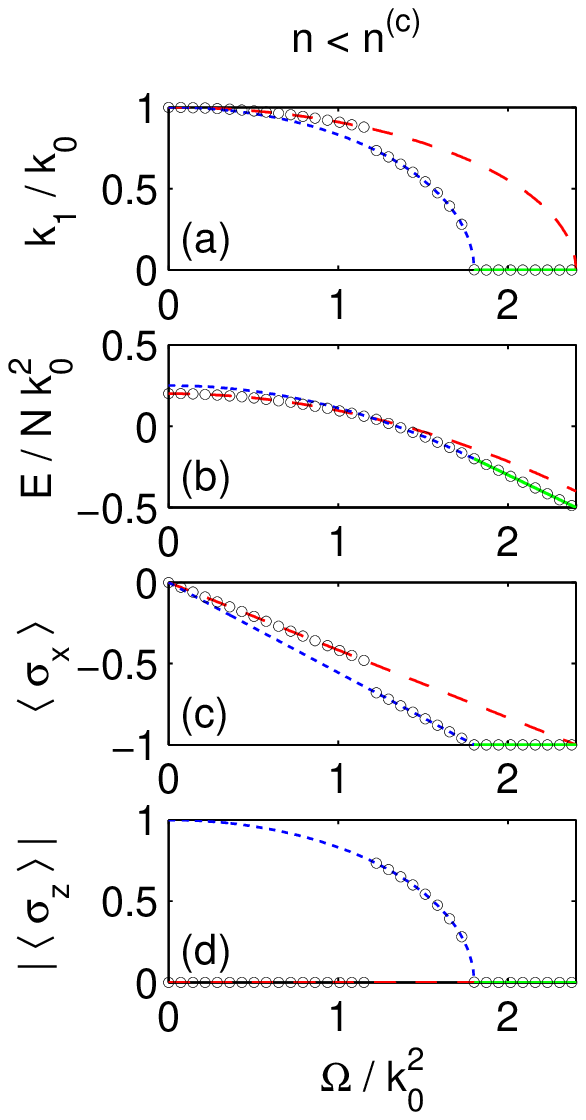}
\hspace{-4mm}\includegraphics[scale=0.678]{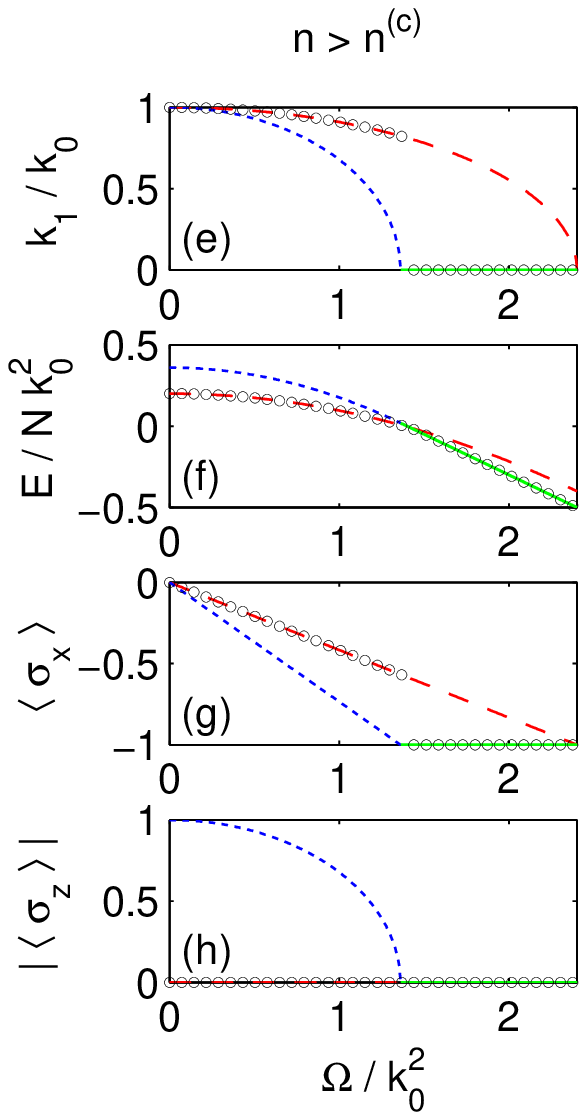}
\caption{(Color online) $k_1$, energy per particle $E/N$, transverse
and longitudinal spin polarization $\langle\sigma_x \rangle$ and
$|\langle \sigma_z\rangle|$ as a function of $\Omega$. Red dashed
lines: stripe phase $k_1\neq 0$ and $\beta= 1/4$; blue dotted lines:
separated phase $k_1\neq 0$ and $\beta=0$; green solid lines: zero
momentum phase $k_1=0$; open circles: ground state. The parameters:
$G_1/k_0^2 = 0.2$, $G_2/k_0^2 = 0.05$ (a)-(d), $G_2/k_0^2 = 0.16$
(e)-(h). } \label{fig:k1_uni_int}
\end{figure}

In Fig.\ref{fig:k1_uni_int}, we plot the momentum $k_1$, the energy
per particle $E/N$, the transverse and longitudinal spin
polarizations $\langle \sigma_x \rangle$ and $|\langle
\sigma_z\rangle|$ as a function of $\Omega$ for $n<n^{(c)}$ (left
column) and for $n>n^{(c)}$ (right column). In addition to the
results for the ground state (open circles), we also show the
various quantities for the three phases (colored lines). Figures
(a)-(d) reveals the emergence of the phase transitions (I-II) and
(II-III), while in (e)-(h) there is only the transition (I-III). The
figures also show that the transitions (I-II) and (I-III) are
accompanied by a jump in $k_1$ [see (a) and (e)] and consequently in
$\langle \sigma_x\rangle$ [see (c) and (g)]. In particular the jump
in $k_1$ associated with the transition (I-III) is sizable and
should be easily observable in experiments. On the other hand only
the transition (I-II) is accompanied by a jump in the longitudinal
spin polarization $|\langle \sigma_z \rangle|$. The transition
(II-III) is instead characterized by a continuous behavior of the
relevant physical parameters. The experimental conditions of
\cite{Lin2011} correspond to values of the average density $n$ much
smaller than $n^{(c)}$, so the jump in $k_1$ could not be detected
because it is too small at the transition (I-II). On the other hand
the occurrence of this phase transition was clearly revealed by the
analysis of the spin distribution after time of flight (see Fig.2c
of \cite{Lin2011}).

\begin{figure}
\includegraphics[scale=0.57]{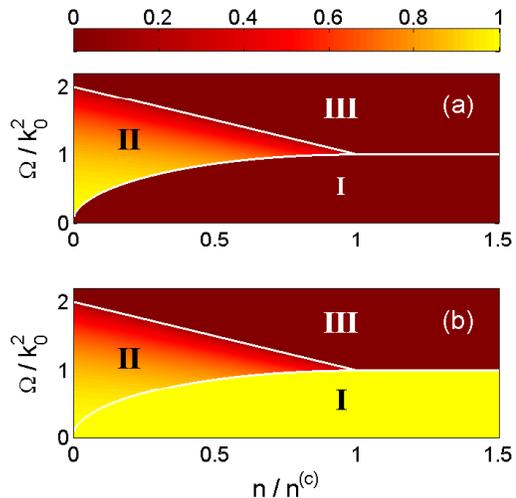}
\caption{(Color online) Spin polarization $|\langle \sigma_z
\rangle|$ (a) and $k_1/k_0$ (b) as a function of $\Omega$ and
density $n/n^{(c)}$ in three different phases with $G_2>0$. The
white solid lines represent the phase transition (I-II), (II-III)
and (I-III). The parameters: $g=100 \,a_B$, where $a_B$ is the Bohr
radius, $\gamma=0.0012$, $k_0^2 = 2\pi\times 80\,$Hz, corresponding
to $n^{(c)}= 4.37\times 10^{15}\,$cm$^{-3}$.}
\label{fig:phase_diagram}
\end{figure}

In Fig.\ref{fig:phase_diagram} we show the phase diagram for the
three different phases. The value of the spin polarization
$|\langle\sigma_z\rangle|$ and $k_1$ are reported in (a) and (b)
respectively. The transition lines separating different phases merge
at a tricritical point at $n=n^{(c)}$. The value of $|\langle
\sigma_z \rangle|$ always vanishes for $n>n^{(c)}$. However the
phase transition (I-III) is well identified by the behavior of the
momentum $k_1$. The parameters employed in
Fig.\ref{fig:phase_diagram} correspond to rather large values of the
critical density. More accessible values of $n^{(c)}$ can be
obtained employing smaller values of $k_0$ or larger values of
$\gamma$ using different spin states or different atomic species.
Reducing the value of $k_0$ would also have the advantage of
increasing the spatial separation between the fringes in the stripe
phase (I), thereby making their experimental detection easier.

The description of the quantum phases carried out in the present
work is based on the mean-field picture which ignores the role of
quantum fluctuations. In ordinary Bose-Einstein condensed gases the
mean-field approach is justified if the gas parameter $n a^3$ is
small. The spin-orbit term in the single-particle Hamiltonian
(\ref{eq:h_0}) is expected to emphasize the role of quantum
fluctuations. In particular when the phase (III) approaches the
phase (II), quantum fluctuations are enhanced and, for large values
of $k_0$, the usual Bogoliubov $\sqrt{n a^3}$ dependence of the
quantum depletion of the condensate is increased by the factor
$(k_0^2/g n)^{1/4}$. The effect is however small for the current
values of the spin-orbit parameters.

Let us now discuss the effect of the trap. In order to simplify the
analysis we have considered harmonic trapping with frequency
$\omega_0$ only along the $x$-axis.  Without interaction, one can
calculate the ground state using a similar variation ansatz,
replacing the plane waves in (\ref{eq:psi_ideal}) by the functions
$e^{\pm i k_1 x} e^{-\omega_0 x^2/2}$, corresponding, in the absence
of the gauge field, to the usual harmonic oscillator Gaussians. The
energy per particle is easily calculated and reads:
\begin{equation}
\begin{aligned}
\varepsilon =& \;\frac{\omega_0}{2} +\frac{k_0^2-k_1^2}{2}-
\frac{\Omega}{2k_0} \sqrt{k_0^2-k_1^2} \\
&-\, \left(C_1^\ast C_2+ C_2^\ast C_1\right) \Omega \,\frac{k_1^2}{2
k_0^2}\, e^{-k_1^2/\omega_0} \; .
\end{aligned}\label{eq:varepsilon_trap}
\end{equation}
The ground state can be found by minimizing $\varepsilon$ with
respect to $k_1$, $C_1$ and $C_2$ with the normalization constraint.
The first term in (\ref{eq:varepsilon_trap}) is just the zero point
energy due to the presence of the trap. The following two terms are
the same as for the uniform case without interactions, i.e.,
(\ref{eq:varepsilon}) with $G_1=G_2=0$. The last term shows the
effect of the trap, fixing the relative phase between the
coefficients $C_1$ and $C_2$ in the ground state. Consequently the
degeneracy occurring in the uniform case will be lifted even in the
absence of interactions (where $\phi=0$). Physically this is the
consequence of the non orthogonality of the two Gaussian states.
According to (\ref{eq:varepsilon_trap}), for $k_1\neq 0$, the system
prefers to stay in the spin mixed phase, and exhibits density
modulation in space even without interactions. On the other hand,
the interaction is crucial for the appearance of the phase separated
configuration. Since the last term of (\ref{eq:varepsilon_trap})
scales exponentially, the effect of the trap is weak for $k_1^2\gg
\omega_0$, and becomes more and more important when $k_1^2$ is
comparable to $\omega_0$.

\begin{figure}
\centering
\includegraphics[scale=0.48]{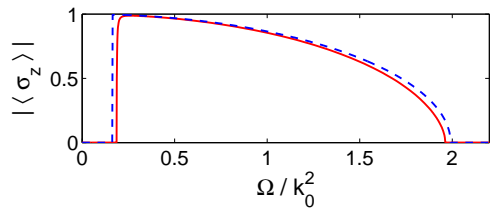}
\caption{(Color online) Spin polarization $|\langle \sigma_z
\rangle|$ as a function of $\Omega$ for the trapped case (red solid
line), and for the uniform case using the density in the center of
the trap (blue dashed line). The parameters are chosen as follows:
$\omega_0=2\pi\times20\,$Hz, $k_0^2/\omega_0 = 4$, $g_{aa}=
g_{bb}=101.20 \,a_B$, $g_{ab} = 100.99\,a_B $, where $a_B$ is the
Bohr radius. The density in the center of trap corresponds to $n
\simeq 3.9\times 10^{13} \, $cm$^{-3}$. } \label{fig:comp_uni_trap}
\end{figure}

To describe the role of the interaction we implement the mean-field
approximation by solving numerically the Gross-Pitaevskii equation
for the condensate wave function using the gradient method in the
same 1D trapping conditions. We find that the properties discussed
in the first part of the work for the uniform system almost hold in
the trapped case. In Fig.\ref{fig:comp_uni_trap} we show an example
of the numerical calculation. The spin polarization as a function of
$\Omega$, in the presence of trapping (red solid line), is compared
with our analytical results for the uniform case (blue dashed line),
using the density in the center of the trap. There is good agreement
between the two curves. We have checked that a similar good
agreement is ensured also for larger values of the interaction
parameter $n/n^{(c)}$, confirming the general validity of the ansatz
(\ref{eq:psi_ideal}) for the spinor wave function employed in the
first part of the Letter.

We finally discuss the case $\delta\neq 0$ and $g_{aa}\neq g_{bb}$,
corresponding to broken spin symmetry. In general one can introduce
three interaction parameters: $G_1= n ( g_{aa}+ g_{bb}+2 g_{ab})/8$,
$G_2= n (g_{aa}+ g_{bb}-2 g_{ab})/8$, and $G_3 = n (g_{aa}
-g_{bb})/4$. In the case of $^{87}$Rb atoms, the scattering lengths
relative to the spin states $|F=1,\,m_F=0 \rangle$ and
$|F=1,\,m_F=-1 \rangle$ are usually parameterized as $a_{aa}\equiv
c_0$, $a_{bb} =c_0+c_2 = a_{ab}$, with $c_0=7.79\times 10^{-12}
\,$Hz\,cm$^3$ and $c_2=-3.61\times10^{-14}\,$Hz\,cm$^3$. This
corresponds to $0<G_2 =G_3 \ll G_1$. However, since the differences
among the scattering lengths are very small, by properly choosing
the detuning $\delta$, this effect can be well compensated, and the
properties of the ground state remain the same as for the spin
symmetric case. For example, using first order perturbation theory,
one finds that correction to the energy per particle is
\begin{equation}
\varepsilon^{(1)}=\left(G_3+\frac{\delta}{2}\right)\frac{k_1}{k_0}
\left(|C_1|^2-|C_2|^2\right) \label{eq:varepsilon_1order}
\end{equation}
where we have considered the low density (weak coupling) limit. By
choosing $\delta= -2 G_3$ the correction
(\ref{eq:varepsilon_1order}) identically vanishes and the transition
frequencies are not consequently affected by the inclusion of the
new terms in the Hamiltonian. Using the $^{87}$Rb parameters
introduced above we find the value
$\Omega^{(\text{I-II})}_{\text{LD}} = 0.19\,E_L$ ($E_L=k^2_0/2$) in
agreement with the findings of \cite{Lin2011} corresponding to
$n/n^{(c)}\ll 1$. For higher densities, the value of $\delta$ should
depend on $\Omega$ dependent in order to ensure exact compensation.

In conclusion, we have investigated the phase diagram of spin-orbit
coupled two-component Bose-Einstein condensates using a variation
ansatz based on the mean-field approximation. We predict a rich
phase diagram characterized by the occurrence of three different
quantum phases, and by a characteristic tricritical point where the
three phases merge at a critical value of the density and of the
Raman frequency. Important questions that remain to be investigated
are the dynamic properties of the system and its behavior at a
finite temperature.

\acknowledgments

Useful discussions with I. Spielman and H. Zhai are acknowledged.
This work has been supported by ERC through the QGBE grant and by
the Italian MIUR through the PRIN-2009 grant.


\begin{thebibliography}{99}

\bibitem{Dalibard2010}
J. Dalibard, F. Gerbier, G. Juzeli\ifmmode \bar{u}\else
\={u}\fi{}nas, and P. \"Ohberg, Rev. Mod. Phys. {\bf 83}, 1523
(2011).

\bibitem{Lin2009}
Y.-J. Lin, R. L. Compton, K. Jim\'{e}nez-Garc\'{i}a, J. V. Porto,
and I. B. Spielman, Nature, {\bf 462}, 628 (2009); Y.-J. Lin, R. L.
Compton, K. Jim\'{e}nez-Garc\'{i}a, W. D. Phillips, J. V. Porto, and
I. B. Spielman, Nature Phys. {\bf 7}, 531 (2011).

\bibitem{Lin2011}
Y.-J. Lin, K. Jim\'{e}nez-Garc\'{i}a, and I. B. Spielman, Nature,
{\bf 471}, 83 (2011).

\bibitem{Hasan2010}
M. Z. Hasan and C. L. Kane, Rev. Mod. Phys., {\bf 82}, 3045 (2010).

\bibitem{Wang2010}
T. D. Stanescu, B. Anderson, and V. Galitski, Phys. Rev. A, {\bf
78}, 023616 (2008); C. Wang, C. Gao, C.-M. Jian, and H. Zhai, Phys.
Rev. Lett., {\bf 105}, 160403 (2010); C.-J. Wu, I. Mondragon-Shem
and X.-F. Zhou, Chin. Phys. Lett., {\bf 28}, 097102 (2011).

\bibitem{Vyasanakere2011}
J. P. Vyasanakere and V. B. Shenoy, Phys. Rev. B, {\bf 83}, 094515
(2011); J. P. Vyasanakere, S. Zhang, and V. B. Shenoy, Phys. Rev. B,
{\bf 84}, 014512 (2011).

\bibitem{Gong2011}
M. Gong, S. Tewari, and C. Zhang, Phys. Rev. Lett., {\bf 107} 195303
(2011); H. Hu, L. Jiang, X.-J. Liu, and H. Pu, Phys. Rev. Lett.,
{\bf 107}, 195304 (2011); Z.-Q. Yu and H. Zhai, Phys. Rev. Lett.,
{\bf 107}, 195305 (2011).

\bibitem{Ho2011}
T.-L. Ho and S. Zhang, Phys. Rev. Lett., {\bf 107}, 150403 (2011).

\bibitem{Rashba1984}
Y. A. Bychkov and E. I. Rashba, J. Phys. C, {\bf 17}, 6039 (1984).

\bibitem{Dresselhaus1955}
G. Dresselhaus, Phys. Rev., {\bf 100}, 580 (1955).

\bibitem{Zhang2011}
Y. Zhang, G. Chen, and C. Zhang, arXiv:1111.4778.



\end{thebibliography}
\end{document}